\documentclass[journal,twoside]{IEEEtran}
\usepackage[cmex10]{amsmath}
\usepackage{amssymb}
\usepackage{url}
\usepackage{multirow}

\newcommand{\F}{{\mathbb F}}
\newcommand{\ftwo}{{\mathbb F}_{2}}
\newcommand{\ftwon}{{\mathbb F}_{2^n}}
\newcommand{\ftwom}{{\mathbb F}_{2^m}}
\newcommand{\ftwok}{{\mathbb F}_{2^k}}

\newcommand{\Tr}{{\operatorname{Tr}}}

\newtheorem{theorem}{Theorem}
\newtheorem{proposition}{Proposition}

\newtheorem{lemma}{Lemma}
\newtheorem{definition}{Definition}

\newtheorem{note}{Note}

\begin{document}

\title{Univariate Niho Bent Functions from o-Polynomials}

\author{Lilya~Budaghyan, Alexander~Kholosha, Claude~Carlet, and Tor~Helleseth,~\IEEEmembership{Fellow,~IEEE}%
\thanks{L. Budaghyan, T. Helleseth, and A. Kholosha are with the Department of Informatics,
University of Bergen, P. O. Box 7803, N-5020 Bergen, Norway (e-mail:
Lilya.Budaghyan@uib.no; Tor.Helleseth@uib.no; Alexander.Kholosha@uib.no).}%
\thanks{C. Carlet is with LAGA, UMR 7539, CNRS, Department of Mathematics,
University of Paris 8 and University of Paris 13, 2 rue de la libert\'e,
93526 Saint-Denis Cedex, France (e-mail: claude.carlet@univ-paris8.fr).}}

\maketitle

\begin{abstract}
In this paper, we discover that any univariate Niho bent function is a sum of
functions having the form of Leander-Kholosha bent functions with extra
coefficients of the power terms. This allows immediately, knowing the terms
of an o-polynomial, to obtain the powers of the additive terms in the
polynomial representing corresponding bent function. However, the
coefficients are calculated ambiguously. The explicit form is given for the
bent functions obtained from quadratic and cubic o-polynomials. We also
calculate the algebraic degree of any bent function in the Leander-Kholosha
class.
\end{abstract}

\begin{IEEEkeywords}
Bent function, Boolean function, maximum nonlinearity, Niho bent function,
o-polynomial, Walsh transform.
\end{IEEEkeywords}

\section{Introduction}
 \label{sec:intro}
Boolean functions of $n$ variables are binary functions over the Galois field
$\ftwon$ (or over the vector space $\ftwo^n$ of all binary vectors of length
$n$). In this paper, we shall always endow this vector space with the
structure of a field, thanks to the choice of a basis of $\ftwon$ over
$\ftwo$. Boolean functions are used in the pseudo-random generators of stream
ciphers and play a central role in their security.

Bent functions were introduced by Rothaus \cite{Ro76} in 1976. These are
Boolean functions of even number of variables $n$, that are maximally
nonlinear in the sense that their Hamming distance to all affine functions is
optimal. This corresponds to the fact that their Walsh transform takes
precisely the values $\pm 2^{n/2}$. Bent functions have also attracted a lot
of research interest because of their relations to coding theory, sequences,
and applications in cryptography. Despite their simple and natural
definition, bent functions turned out to admit a very complicated structure
in general. On the other hand, many special explicit constructions are known.
Distinguished are primary constructions giving bent functions from scratch
and secondary ones building new bent functions from one or several given bent
functions.

Bent functions are often better viewed in their bivariate representation but
can also be viewed in their univariate form (see Section~\ref{sec:prelim}). A
good survey reference containing information on explicit primary
constructions of bent functions in their univariate form (expressed by means
of the trace function) is \cite{Ca10,KhPo13}. It is well known that some of
these explicit constructions belong to the two general families of bent
functions which are the original Maiorana-McFarland \cite{McFa73} and the
Partial Spreads ($\mathcal{PS}$) classes. It was in the early 1970s when
Dillon in his thesis \cite{Di74} introduced the two above mentioned classes
and also another one denoted by $H$, where bentness is proven under some
conditions which were not obvious to achieve (in this class, Dillon was able
to exhibit only functions belonging, up to the affine equivalence, to the
Maiorana-McFarland class).

It was observed in \cite{CaMe11} that the class of the, so called, Niho bent
functions (introduced in \cite{DoLeCaCaFeGa06} by  Dobbertin \emph{et al})
is, up to EA-equivalence, equal to the Dillon's class $H$. Note that
functions in class $H$ are defined in their bivariate representation and Niho
bent functions had originally a univariate form only. Three infinite families
of Niho binomial bent functions were constructed in \cite{DoLeCaCaFeGa06} and
one of these constructions was later generalized by Leander and Kholosha
\cite{LeKh06} into a function with $2^r$ Niho exponents. Another class was
also extended in \cite{HeKhMe12}. In \cite{BuCaHeKhMe12} it was proven that
some of these infinite families of Niho bent functions are EA-inequivalent to
any Maiorana-McFarland function which implies that classes $H$ and
Maiorana-McFarland are different up to EA-equivalence. New classes of Niho
bent functions were also introduced in \cite{CaMe11} thanks to the observed
connection between class $H$ and o-polynomials.

In this paper, we prove that any univariate Niho bent function is a sum of
functions having the form of Leander-Kholosha bent function (see
\cite{LeKh06}) with extra coefficients of the power terms. In particular, any
o-monomial corresponds to a $2^r$ term Niho bent function of Leander-Kholosha
type with coefficients of the power terms inserted. This result allows
immediately, knowing the terms of an o-polynomial, to obtain the powers of
the additive terms in the polynomial representing corresponding bent
function. However, the coefficients are calculated ambiguously. The explicit
form is given for the bent functions obtained from quadratic and cubic
o-polynomials. In general, we provide an explicit form for all Niho bent
functions that correspond to o-monomials and o-polynomials of degree two and
three. We also succeed in calculating the algebraic degree of any bent
function in the Leander-Kholosha class. The paper is organized as follows. In
Section~\ref{sec:prelim}, we fix our main notation, recall the necessary
background and, in Subsection~\ref{sssec:Niho} study the algebraic degree.
Further, in Section~\ref{sec:classH}, we describe briefly the class
$\mathcal{H}$ introduced in \cite{CaMe11} and give some necessary facts that
we need later. The quadratic and cubic o-polynomials and their corresponding
bent functions are considered in
Sections~\ref{sec:new_qu}~and~\ref{sec:new_cu}.

\section{Notation and Preliminaries}
 \label{sec:prelim}
For any set $E$, denote $E\setminus\{0\}$ by $E^*$. Throughout the
paper, let $n$ be even and $n=2m$.

\subsection{Trace Representation, Boolean Functions in Univariate
and Bivariate Forms} For any positive integer $k$ and any $r$ dividing $k$,
the trace function $\Tr^k_r()$ is the mapping from $\ftwok$ to $\F_{2^r}$
defined by
\[\Tr^k_r(x):=\sum_{i=0}^{\frac{k}{r}-1}x^{2^{ir}}=x+x^{2^r}+x^{2^{2r}}+\cdots+x^{2^{k-r}}\enspace.\]
In particular, the {\em absolute trace} over $\ftwok$ is the function
$\Tr^k_1(x)=\sum_{i=0}^{k-1}x^{2^i}$ (in what follows, we just use $\Tr_k()$
to denote the absolute trace). Recall that the trace function satisfies the
transitivity property $\Tr_k=\Tr_r\circ\Tr^k_r$.

The univariate representation of a Boolean function is defined as follows: we
identify $\ftwo^n$ (the $n$-dimensional vector space over $\ftwo$) with
$\ftwon$ and consider the arguments of $f$ as elements in $\ftwon$. An inner
product in $\ftwon$ is $x\cdot y=\Tr_n(xy)$. There exists a unique univariate
polynomial $\sum_{i=0}^{2^n-1}a_i x^i$ over $\ftwon$ that represents $f$
(this is true for any vectorial function from $\ftwon$ to itself). The
algebraic degree of $f$ is equal to the maximum $2$-weight of an exponent
having nonzero coefficient, where the $2$-weight $w_2(i)$ of an integer $i$
is the number of ones in its binary expansion. Moreover, $f$ being Boolean,
its univariate representation can be written uniquely in the form of
\[f(x)=\sum_{j\in\Gamma_n}\Tr_{o(j)}(a_j x^j)+a_{2^n-1} x^{2^n-1}\enspace,\]
where $\Gamma_n$ is the set of integers obtained by choosing the smallest
element in each cyclotomic coset modulo $2^n-1$ (with respect to $2$), $o(j)$
is the size of the cyclotomic coset containing $j$, $a_j\in\F_{2^{o(j)}}$ and
$a_{2^n-1}\in\ftwo$. The function $f$ can also be written in a non-unique way
as $\Tr_n(P(x))$ where $P(x)$ is a polynomial over $\ftwon$.

The bivariate representation of a Boolean function is defined as follows: we
identify $\ftwo^n$ with $\ftwom\times\ftwom$ and consider the argument of $f$
as an ordered pair $(x,y)$ of elements in $\ftwom$. There exists a unique
bivariate polynomial $\sum_{0\leq i,j\leq 2^m-1}a_{i,j}x^i y^j$ over $\ftwom$
that represents $f$. The algebraic degree of $f$ is equal to
$\max_{(i,j)\,|\,a_{i,j}\neq 0}(w_2(i)+w_2(j))$. And $f$ being Boolean, its
bivariate representation can be written in the form $f(x,y)=\Tr_m(P(x,y))$,
where $P(x,y)$ is some polynomial of two variables over $\ftwom$.

\subsection{Walsh Transform and Bent Functions}
Let $f$ be an $n$-variable Boolean function. Its {\em ``sign" function} is
the integer-valued function $\chi_f:=(-1)^f$. The {\em Walsh transform} of
$f$ is the discrete Fourier transform of $\chi_f$ whose value at point
$w\in\ftwon$ is defined by
\[\hat\chi_f(w)=\sum_{x\in\ftwon}(-1)^{f(x)+\Tr_n(wx)}\enspace.\]

\begin{definition}
For even $n$, a Boolean function $f$ in $n$ variables is said to be {\em
bent} if for any $w\in\ftwon$ we have $\hat\chi_f(w)=\pm 2^{\frac{n}{2}}$.
\end{definition}

It is well known (see, for instance, \cite{Ca10}) that the algebraic
degree of a bent Boolean function in $n>2$ variables is at most
$\frac{n}{2}$. This means that in the univariate representation of a
bent function, all exponents $i$ whose $2$-weight is larger than $m$
have zero coefficients $a_i$. If $f$ is a bent function in $n$
variables then its dual $\tilde{f}$ is the Boolean function defined
by
\[\hat\chi_f(w)=2^{\frac{n}{2}}\chi_{\tilde{f}}(w)\enspace.\]
Obviously, $\tilde{f}$ is also bent and its dual is $f$ itself.

\begin{definition}
Functions $f,g:\ftwo^n\to\ftwo$ are {\em extended-affine equivalent} (in
brief, EA-equivalent) if there exist affine permutation $L$ of $\ftwo^n$ and
an affine function $l:\ftwo^n\to\ftwo$ such that $g(x)=(f\circ L)(x)+l(x)$. A
class of functions is {\em complete} if it is a union of EA-equivalence
classes. The {\em completed class} is the smallest possible complete class
that contains the original one.
\end{definition}

\subsection{Niho Power Functions}
 \label{sssec:Niho}
A positive integer $d$ (always understood modulo $2^n-1$ with $n=2m$) is a
{\em Niho exponent} and $t\to t^d$ is a {\em Niho power function} if the
restriction of $t^d$ to $\ftwom$ is linear or, equivalently, $d\equiv 2^j\
\pmod{2^m-1}$ for some $j<n$. As we consider $\Tr_n(at^d)$ with $a\in\ftwon$,
without loss of generality, we can assume that $d$ is in the normalized form,
i.e., with $j=0$. Then we have a unique representation $d=(2^m-1)s+1$ with
$1<s<2^m+1$. If some $s$ is written as a fraction, this has to be interpreted
modulo $2^m+1$ (e.g., $1/2=2^{m-1}+1$). Following are examples of bent
functions consisting of one or more Niho exponents:
\renewcommand{\theenumi}{\arabic{enumi}}
\renewcommand{\labelenumi}{\theenumi.}
\begin{enumerate}
\item\label{it:quad} Quadratic function $\Tr_m(at^{2^m+1})$ with
    $a\in\ftwom^*$ (here $s=2^{m-1}+1$).
\item Binomials of the form $f(t)=\Tr_n(\alpha_1 t^{d_1}+\alpha_2
    t^{d_2})$, where $2d_1\equiv 2^m+1\pmod{2^n-1}$ and
    $\alpha_1,\alpha_2\in\ftwon^*$ are such that
    $(\alpha_1+\alpha_1^{2^m})^2=\alpha_2^{2^m+1}$. Equivalently,
    denoting $a=(\alpha_1+\alpha_1^{2^m})^2$ and $b=\alpha_2$ we have
    $a=b^{2^m+1}\in\ftwom^*$ and
    \[f(t)=\Tr_m(at^{2^m+1})+\Tr_n(bt^{d_2}).\]
    We note that if $b=0$ and $a\neq 0$ then $f$ is a bent function
    listed under number~\ref{it:quad}. The possible values of $d_2$ are
    \cite{DoLeCaCaFeGa06,HeKhMe12}:
\[\begin{split}
    d_2&=(2^m-1)3+1,\\
    6d_2&=(2^m-1)+6\mbox{ (taking $m$ even)}.
\end{split}\]
    These functions have algebraic degree $m$ and do not belong to the
    completed Maiorana-McFarland class \cite{BuCaHeKhMe12}.
\item \cite{LeKh06,LiHeKhTa13} Take $1<r<m$ with $\gcd(r,m)=1$ and define
    \begin{equation}
     \label{eq:LK1}
    f(t)=\Tr_n\Bigg(a^2t^{2^m+1}+(a+a^{2^m})\sum_{i=1}^{2^{r-1}-1}t^{d_i}\Bigg),
    \end{equation}
    where $2^r d_i=(2^m-1)i+2^r$ and $a\in\ftwon$ is such that
    $a+a^{2^m}\neq 0$. This function has algebraic degree $r+1$ (see
    Proposition~\ref{pr:deg}) and belongs to the completed
    Maiorana-McFarland class \cite{CaHeKhMe11}. On the other hand, the
    dual of $f$ is not a Niho bent function \cite{CaHeKhMe11}.
\item Bent functions in a bivariate representation obtained from the
    known o-polynomials.
\end{enumerate}

Consider the listed above two binomial bent functions. If $\gcd(d_2,2^n-1)=d$
and $b=\beta^d$ for some $\beta\in\ftwon$ then $b$ can be ``absorbed" in the
power term $t^{d_2}$ by a linear substitution of variable $t$. In this case,
up to EA-equivalence, $b=a=1$. In particular, this applies to any $b$ when
$\gcd(d_2,2^n-1)=1$ that holds in both cases except when $d_2=(2^m-1)3+1$
with $m\equiv 2\pmod{4}$ where $d=5$. In this exceptional case, we can get up
to three different classes (since exponents $1$, $2$ and $4$ belong to the
same cyclotomic coset) but the exact situation has to be further
investigated.

Also, it can be easily seen that in function (\ref{eq:LK1}), up to
EA-equivalence, we can assume $a+a^{2^m}=1$. Indeed, let
$a+a^{2^m}=b\in\ftwom$ and substitute $t$ in (\ref{eq:LK1}) for $b^{-1}t$.
This results in a function having the same form as $f(t)$ except for $a/b$
taken instead of $a$. It remains to note that $a/b+(a/b)^{2^m}=1$. Also note
that the conjugated exponent $d_i$ is equal to
\[2^m((2^m-1)i2^{-r}+1)=(2^m-1)(2^{m-r}i+1)+1\]
and, therefore, bent function (\ref{eq:LK1}) can be equivalently written as
\begin{equation}
 \label{eq:LK2}
\Tr_n\bigg(a^2t^{2^m+1}+(a+a^{2^m})\sum_{i=1}^{2^{r-1}-1}t^{(2^m-1)(2^{m-r}i+1)+1}\bigg).
\end{equation}
We will use this representation when extending this class in the following
sections.

\begin{proposition}
 \label{pr:deg}
Function $f(t)$ in (\ref{eq:LK2}) has algebraic degree $r+1$.
\end{proposition}

\begin{IEEEproof}
For any $i\in\{1,\dots,2^{r-1}-1\}$ take exponent $(2^{m-r}i+1)(2^m-1)+1$ and
analyze its binary expansion.

First, for any odd $l=\sum_{i=0}^{m-1}l_i 2^i$ being its binary expansion, we
obtain
\[\begin{split}
l(2^m-1)&=\sum_{i=1}^{m-1}l_i 2^{m+i}+l_0 2^m-\sum_{i=0}^{m-1}l_i 2^i\\
&=\sum_{i=1}^{m-1}l_i 2^{m+i}+(l_0-1)2^m+1+\sum_{i=0}^{m-1}(1-l_i)2^i\\
&=\sum_{i=1}^{m-1}l_i 2^{m+i}+1+\sum_{i=1}^{m-1}(1-l_i)2^i
\end{split}\]
since $l_0=1$. Therefore, for $l=2^{m-r}i+1$ we obtain
\[\begin{split}
{\rm wt}(l(2^m-1)+1)&={\rm wt}(2^m-2-(l-1)+2)+{\rm wt}(l)-1\\
&={\rm wt}(2^m-2^{m-r}i)+{\rm wt}(i)\\
&={\rm wt}(2^r-i)+{\rm wt}(i)\\
&=r-{\rm wt}(i-1)+{\rm wt}(i)\\
&=r-s+1\enspace,
\end{split}\]
where $i=2^s j$ with $s\geq 0$ and $j$ odd.

Thus, the maximal weight of exponents in $f(t)$ is $r+1$. We complete the
proof by showing that all the exponents in (\ref{eq:LK2}) are cyclotomic
inequivalent. Assume, on the contrary, there exist
$i,j\in\{1,\dots,2^{r-1}-1\}$ with $i\neq j$ and $t\in\{0,\dots,2m-1\}$ such
that
\[\begin{split}
&2^{m-r}i(2^m-1)+2^m\equiv 2^t\big(2^{m-r}j(2^m-1)+2^m\big)\ \mbox{or}\\
&2^{m-r}(2^m-1)(2^t j-i)+2^m(2^t-1)\equiv 0\pmod{2^{2m}-1}
\end{split}\]
that holds only if $2^m-1$ divides $2^t-1$ that gives $t=m$ (for $t=0$,
obviously, $i=j$). This results in the following equivalence
\[2^{m-r}(2^m j-i)+2^m\equiv 0\pmod{2^m+1}\]
that has a unique solution $i=2^r-j$ modulo $2^m+1$. These solutions are not
good since we have that $0<i<2^{r-1}$.
\end{IEEEproof}

Note that bent function is obtained in (\ref{eq:LK1}) also when $r>m+1$.
However, both $r$ and $r-m$ in this case result in bent functions that are
the same, up to addition of a linear term. Indeed, assume $r=m+s$ with
$1<s<m$ and $\gcd(s,m)=1$. Then, after multiplying $d_i$ by $2^{2m}$ (that is
one modulo $2^n-1$) we obtain
\[d_i=(2^m-1)2^{m-s}i+1\quad\mbox{for}\quad i=1,\dots,2^{m+s-1}-1\enspace.\]
Since $i$ can be reduced modulo $2^m+1$, the last $(2^m+1)(2^{s-1}-2)$ power
terms in (\ref{eq:LK1}) cancel out and we are left with the terms
corresponding to $i=1,\dots,2^{m+1}-2^{s-1}+1$. For the same reason, more
terms cancel out that shrinks the range to $i=2^m-2^{s-1}+1,\dots,2^m+1$.
Further, taking $i=2^m-2^{s-1}+1$ we get
\[d_i=-2^{2m-1}+2^{m-1}+1\equiv 2^{m-1}(2^m+1)\pmod{2^{2m}-1}\]
and $\Tr^n_m(t^{d_i})=0$ since $t^{d_i}\in\ftwom$. Also, taking $i=2^m+1$ we
get $d_i\equiv 1\pmod{2^{2m}-1}$ that gives a linear term.

The remaining $2^{s-1}-1$ terms correspond to $i=2^m-2^{s-1}+2,\dots,2^m$.
Taking $i=2^m-2^{s-1}+2$ we obtain that
\[\begin{split}
2^{m-s}i&=(2^{m-s}-1)(2^m+1)+2^{m-1}+2^{m-s}+1\\
&\equiv 2^{m-1}+2^{m-s}+1\pmod{2^m+1}\enspace.
\end{split}\]
Therefore,
\[d_i=(2^m-1)(2^{m-1}+2^{m-s}i+1)+1\ \mbox{for}\ i=1,\dots,2^{s-1}-1\enspace.\]
Finally,
\[\begin{split}
2^m d_i&=(2^m-1)(2^{2m-1}+2^{2m-s}i+2^m+1)+1\\
&\equiv (2^m-1)(2^{m-1}-2^{m-s}i+1)+1\\
&= (2^m-1)\big(2^{m-s}(2^{s-1}-i)+1\big)+1\pmod{2^n-1}
\end{split}\]
which indicates that $d_i$ are $2^m$th powers of the exponents in
(\ref{eq:LK2}) taken with $r=s$. Also raising to the power of $2^m$ does not
change the coefficient $a+a^{2^m}$.

Consider the remaining case when $r=m+1$ and
\[d_i=(2^m-1)2^{m-1}i+1\quad\mbox{for}\quad i=1,\dots,2^m-1\enspace.\]
Obviously, for $i<2^m-1$,
\[\begin{split}
2^m d_i&=(2^m-1)(2^{2m-1}i+1)+1\\
&\equiv(2^m-1)(-2^{m-1}i+2^{2m-1}-2^{m-1})+1\\
&=(2^m-1)\big(2^{m-1}(2^m-1-i)\big)+1\pmod{2^n-1}\\
&=d_{2^m-1-i}\enspace.
\end{split}\]
Therefore, all power terms in (\ref{eq:LK2}) cancel out except for the
quadratic one and the one corresponding to $i=2^m-1$ having
\[d_{2^m-1}=(2^m-1)^2 2^{m-1}+1\equiv 2^m\pmod{2^n-1}\]
and we get
\[f(t)=\Tr_n\big(a^2t^{2^m+1}+(a+a^{2^m})t\big)\]
that is, ignoring the linear term, a quadratic bent function listed under
number~\ref{it:quad}.

\section{Class $\mathcal{H}$ of Bent Functions}
 \label{sec:classH}
In his thesis \cite{Di74}, Dillon introduced the class of bent
functions denoted by $H$. The functions in this class are defined in
their bivariate form as
\begin{equation}
 \label{eq:H}
f(x,y)=\Tr_m\big(y+xF(yx^{2^m-2})\big),
\end{equation}
where $x,y\in\ftwom$ and $F$ is a permutation of $\ftwom$ such that $F(x)+x$
does not vanish and for any $\beta\in\ftwom^*$, the function $F(x)+\beta x$
is $2$-to-$1$ (i.e., the pre-image of any element of $\ftwom$ is either a
pair or the empty set). The condition that $F(x)+x$ does not vanish is
required only for (\ref{eq:H}) to belong to $\mathcal{PS}$ but is not
necessary for bentness. Dillon was just able to exhibit bent functions in $H$
that also belong to the completed Maiorana-McFarland class. As observed by
Carlet and Mesnager \cite[Proposition~1]{CaMe11}, this class can be slightly
extended into a class $\mathcal{H}$ defined as the set of (bent) functions
$g$ satisfying
 \setlength{\arraycolsep}{0.14em}
\begin{equation}
 \label{eq:exH}
g(x,y)=\left\{\begin{array}{ll}
\Tr_m\left(xG\left(\frac{y}{x}\right)\right),&\ \mbox{if}\quad x\neq 0\\
\Tr_m(\mu y),&\ \mbox{if}\quad x=0\enspace,
\end{array}\right.
\end{equation}
 \setlength{\arraycolsep}{5pt}\noindent
where $\mu\in\ftwom$ and $G$ is a mapping from $\ftwom$ to itself satisfying
the following necessary and sufficient conditions:
 \setlength{\arraycolsep}{0.14em}
\begin{eqnarray}
 \label{eq:co1}
F:z&\to&G(z)+\mu z\ \mbox{is a permutation on}\ \ftwom\\
 \label{eq:co2}
z&\to&F(z)+\beta z\ \mbox{is 2-to-1 on}\ \ftwom\ \mbox{for any}\ \beta\in\ftwom^*.\quad
\end{eqnarray}
 \setlength{\arraycolsep}{5pt}\noindent
As proved in \cite{CaMe11}, condition (\ref{eq:co2}) implies condition
(\ref{eq:co1}) and, thus, is necessary and sufficient for $g$ being bent.
Adding the linear term $\Tr_m((\mu+1)y)$ to (\ref{eq:exH}) we obtain the
original Dillon function (\ref{eq:H}). Therefore, functions in $\mathcal{H}$
and in the Dillon class are the same up to the addition of a linear term. It
is observed in \cite{CaMe11} that Niho bent functions are just functions in
$\mathcal{H}$ in the univariate representation.

Any mapping $F$ on $\ftwom$ that satisfies (\ref{eq:co2}) is called an
\emph{o-polynomial}. The only linear o-monomial is a Frobenius map
\[F(z)=z^{2^i}\quad\mbox{with}\quad \gcd(i,m)=1\enspace.\]
As proven in \cite{ChSt98}, following is the list of \emph{all existing}
quadratic o-monomials.
\renewcommand{\theenumi}{\arabic{enumi}}
\renewcommand{\labelenumi}{\theenumi.}
\begin{enumerate}
\item $F(z)=z^6$ with $m$ odd.
\item $F(z)=z^{2^{2k}+2^k}$ with $m=4k-1$.
\item $F(z)=z^{2^{3k+1}+2^{2k+1}}$ with $m=4k+1$.
\item $F(z)=z^{2^k+2}$ with $m=2k-1$.
\item\label{it:5} $F(z)=z^{2^{m-1}+2^{m-2}}$ with $m$ odd.
\end{enumerate}
In \cite{Vi10}, it was shown that the only cubic o-monomial is
\[F(z)=z^{3\cdot 2^k+4}\quad\mbox{with}\quad m=2k-1\enspace.\]
It is conjectured that no other o-monomial exists. Further, two o-trinomials
are found
\[\begin{array}{lll}
F(z)=z^{2^k}+z^{2^k+2}+z^{3\cdot 2^k+4}&\mbox{with}&m=2k-1\\
F(z)=z^{\frac{1}{6}}+z^{\frac{1}{2}}+z^{\frac{5}{6}}&\mbox{with}&m\ \mbox{odd}\enspace.
\end{array}\]
The remaining two known, up to equivalence, o-polynomials are Subiaco and
Adelaide listed in \cite{CaMe11}.

Using (\ref{eq:exH}), every o-polynomial results in a bent function in class
$\mathcal{H}$ (and vice versa). In particular, functions (\ref{eq:LK1}) with
$a+a^{2^m}=1$ are obtained from Frobenius map $z^{2^{m-r}}$
\cite{CaHeKhMe11}, binomial Niho bent functions with $d_2=(2^m-1)3+1$
correspond to Subiaco hyperovals \cite{HeKhMe12} and functions with
$6d_2=(2^m-1)+6$ correspond to Adelaide hyperovals. In the following
Section~\ref{sec:new_qu}, we find bent functions that correspond to all the
existing quadratic o-monomials. In Section~\ref{sec:new_cu} the same problem
is resoled for all cubic o-monomials.

\section{General Form of a Niho Bent Function}
 \label{sec:Niho_genf}
By definition, all exponents of monomials contained in the univariate
representation of a Niho bent function are of the Niho type, i.e., have the
form of $d=(2^m-1)s+1$ with $1<s<2^m+1$. From the results in this section, in
particular, it follows that in a Niho bent function, $s$ is odd. Moreover, we
prove that any Niho bent function, up to EA-equivalence, is obtained as a sum
of the following functions
\begin{equation}
 \label{eq:LK_coef}
\Tr_n\bigg(A_{2^{r-1}}t^{2^m+1}+\sum_{i=1}^{2^{r-1}-1}A_i
t^{(2^m-1)(2^{m-r}i+1)+1}\bigg)
\end{equation}
with $0<r<m$ and $A_i\in\ftwon^*$. Each function making up the sum is defined
by a monomial found in the corresponding o-polynomial and has a particular
set of nonzero coefficients $A_i$. Parameter $0<m-r<m$ is equal to the
position of the least significant one-digit in the binary expansion of the
exponent in this monomial. The whole sum also has the form of
(\ref{eq:LK_coef}) (taken with the largest $r$ found among all the additive
components) but some terms may cancel out due to addition of coefficients.
Note that (\ref{eq:LK_coef}) consists of the same power terms as
Leander-Kholosha bent function (\ref{eq:LK2}) but also has a particular
coefficient for each term.

\begin{lemma}
 \label{le:mon}
Take an integer $d\in\{1,\dots,2^m-1\}$ and let $l\in\{0,\dots,m-1\}$ be the
position of the least significant one-digit in the binary expansion of $d$.
Take any $\lambda\in\ftwom^*$ and define bivariate function
$g(x,y)=\Tr_m(\lambda x^{2^m-d}y^d)$ over $\ftwom\times\ftwom$. Then the
univariate form of $g(x,y)$ obtained using identities $x=t+t^{2^m}$ and
$y=at+a^{2^m}t^{2^m}$, where $t\in\ftwon$ and $a$ is a primitive element of
$\ftwon$, has the form of (\ref{eq:LK_coef}) with $m-r=l$, plus a linear
term.
\end{lemma}

\begin{IEEEproof}
Denote $I_k=\{0,\dots,k-1\}$ for $k>0$ and assume $I_0=\emptyset$. Define
$D\subset I_m$ such that $d=\sum_{i\in D} 2^i$. Also define $T\subset I_m$
such that $2^m-d=\sum_{i\in T} 2^i$. It is easy to see that
\[T=\big(I_m\setminus (D\cup I_l)\big)\cup \{l\}\enspace.\]
Note that $D\cap T=\{l\}$ and $D\cup T=I_m\setminus I_l$.

Further,
\begin{align}
 \label{eq:int}
\nonumber &(t+t^{2^m})^{2^m-d}(at+a^{2^m}t^{2^m})^d\\
\nonumber &=\prod_{i\in T}(t^{2^i}+t^{2^{m+i}})\prod_{j\in D}(a^{2^j}t^{2^j}+a^{2^{m+j}}t^{2^{m+j}})\\
\nonumber &=\sum_{c_i\in\{0,1\} \atop i\in T}t^{\sum_{i\in T}(c_i 2^i+\overline{c_i} 2^{m+i})}\\
\nonumber &\ \times\sum_{s_j\in\{0,1\} \atop j\in D}a^{\sum_{j\in D}(s_j 2^j+\overline{s_j} 2^{m+j})}t^{\sum_{j\in D}(s_j 2^j+\overline{s_j} 2^{m+j})}\\
\nonumber &=\sum_{c_i,s_j\in\{0,1\} \atop i\in T,\,j\in D}a^{\sum_{j\in D}(s_j 2^j+\overline{s_j} 2^{m+j})}\\
\nonumber &\ \times t^{\sum_{i\in T}(c_i 2^i+\overline{c_i} 2^{m+i})+\sum_{j\in D}(s_j 2^j+\overline{s_j} 2^{m+j})}\\
\nonumber &=\sum_{c_i,s_l\in\{0,1\} \atop i\in I_m\setminus I_l}a^{\sum_{i\in D\setminus\{l\}}(c_i 2^i+\overline{c_i} 2^{m+i})+s_l 2^l+\overline{s_l} 2^{m+l}}\\
\nonumber &\ \times t^{\sum_{j\in I_m\setminus I_l}(c_j 2^j+\overline{c_j} 2^{m+j})+s_l 2^l+\overline{s_l} 2^{m+l}}\\
&=\sum_{c_i,s\in\{0,1\} \atop i\in I_m\setminus I_l}a^{\sum_{i\in D\setminus\{l\}}(c_i 2^i+\overline{c_i} 2^{m+i})+s 2^l+\overline{s} 2^{m+l}}\\
\nonumber &\ \times t^{2^l c+2^{m+l}(2^{m-l}-c-1)+s 2^l+\overline{s} 2^{m+l}}\enspace,
\end{align}
where integer $c=(c_{m-1},\dots,c_l)$ in its binary expansion with the least
significant bit $c_l$ and the line over a bit value denotes its complement.
Note that $0\leq c<2^{m-l}$.

Now we make several observations on additive terms in (\ref{eq:int}):
\renewcommand{\theenumi}{\roman{enumi}}
\renewcommand{\labelenumi}{(\theenumi)}
\begin{enumerate}
\item Assume $c=2^{m-l}-1$ and $s=1$. Then the corresponding term is
    equal to $a^d t^{2^m}$ since
    \[\sum_{i\in D\setminus\{l\}} 2^i+2^l=d\enspace.\]
\item Take any $c\in\{0,\dots,2^{m-l}-2\}$ and $s=1$. Then the power of
    $t$ in the corresponding term is equal to
    \[\begin{split}
    &2^l c+2^{m+l}(2^{m-l}-c-1)+2^l\\
    &=2^l(c+1)+2^{m+l}(2^{m-l}-c-2)+2^{m+l}
    \end{split}\]
    that is equal to the power of $t$ in the term corresponding to $c+1$
    and $s=0$.

    In particular, taking $c=2^{m-l-1}-1$ with $s=1$ (or $c=2^{m-l-1}$
    with $s=0$) we obtain the same power of $t$ equal to
    \[\begin{split}
    &2^l(2^{m-l-1}-1)+2^{m+l}(2^{m-l}-2^{m-l-1})+2^l\\
    &=2^{m-1}(2^m+1)\enspace.
    \end{split}\]
    This exponent is a self-conjugate. The coefficient of this term is
    equal to $a^{\tilde{d}}+a^{2^m\tilde{d}}$ with
\setlength{\arraycolsep}{0.14em}
    \[\begin{split}
    &\tilde{d}=\sum_{i\in D\setminus\{l\}}(c_i 2^i+\overline{c_i} 2^{m+i})+2^l=\sum_{i\in D}(c_i 2^i+\overline{c_i} 2^{m+i})\\
    &=\left\{\begin{array}{ll}
    d,&\ \mbox{if}\ m-1\notin D\\
    d+2^{m-1}(2^m-1),&\ \mbox{otherwise}
     \end{array}\right.
    \end{split}\]
\setlength{\arraycolsep}{5pt}\noindent since $c=2^{m-l-1}-1$.
\item Take any $c\in\{0,\dots,2^{m-l}-1\}$. The powers of $t$ in the
    terms corresponding to $c$ with $s=1$ and $2^{m-l}-c-1$ with $s=0$
    are conjugates since
    \[\begin{split}
    &\big(2^l c+2^{m+l}(2^{m-l}-c-1)+2^l\big)2^m\\
    &\equiv 2^l(2^{m-l}-c-1)+2^{m+l}c+2^{m+l}\pmod{2^n-1}\enspace.
    \end{split}\]

    It is obvious that the powers of $a$ in the terms corresponding to
    $c$ with $s=1$ and $2^{m-l}-c-1$ with $s=0$ are conjugates as well.
\end{enumerate}

Therefore, we can fix $c_{m-1}=1$, $s=0$ and rewrite (\ref{eq:int}) as
\[\begin{split}
&\Tr_m^n\bigg(a^d t^{2^m}+a^{\tilde{d}} t^{2^{m-1}(2^m+1)}\\
&\ +\sum_{c_i\in\{0,1\};\,c'>0 \atop i\in I_{m-1}\setminus I_l}\Big(a^{\sum_{i\in D\setminus\{l\}\atop c_{m-1}=1}(c_i 2^i+\overline{c_i} 2^{m+i})+2^{m+l}}\\
&\ +a^{\sum_{i\in D\setminus\{l\}\atop c_{m-1}^*=1}(c_i^* 2^i+\overline{c_i^*} 2^{m+i})+2^l}\Big)\\
&\ \times t^{2^l c'+2^{m+l}(2^{m-l-1}-c'-1)+2^{m-1}+2^{m+l}}\bigg)\\
&=\Tr_m^n\bigg(a^d t^{2^m}+a^{\tilde{d}} t^{2^{m-1}(2^m+1)}\\
&\ +\sum_{c'=1}^{2^{m-l-1}-1}A_{c'} t^{(2^m-1)2^l(2^{m-l-1}-c')+2^m}\bigg)\enspace,
\end{split}\]
where $c'=(c_{m-2},\dots,c_l)$ and $c'-1=(c_{m-2}^*,\dots,c_l^*)$ in its
binary expansion with the least significant bit $c_l$ and $A_i\in\ftwon$ are
defined explicitly. In particular, since $a$ is a primitive element of
$\ftwon$, we conclude that all coefficients $A_i$ are nonzero. In the case
when $l=m-1$ the sum over $c_i$ is empty.

Finally, multiplying the latter expression by $\lambda$ and placing it under
the $\Tr_m()$ function, ignoring the linear term $\Tr_n(a^d t^{2^m})$, we
obtain the expression having the form of (\ref{eq:LK_coef}) with $m-r=l$.
\end{IEEEproof}

Observe some important properties of coefficients $A_{c'}$.
\renewcommand{\theenumi}{\roman{enumi}}
\renewcommand{\labelenumi}{(\theenumi)}
\begin{enumerate}
\item  For any $c'\in\{1,\dots,2^{m-l-2}\}$
    \setlength{\arraycolsep}{0.14em}
    \[A_{c'}^{2^m}=\left\{\begin{array}{ll}
    A_{2^{m-l-1}-c'},&\ \mbox{if}\ m-1\notin D\\
    a^{2^{m-1}(2^m-1)}A_{2^{m-l-1}-c'},&\ \mbox{otherwise.}
     \end{array}\right.\]
    \setlength{\arraycolsep}{5pt}\noindent
    Indeed, if $m-1\notin D$ then
\[\begin{split}
A_{c'}^{2^m}&=a^{\sum_{i\in D\setminus\{l\}}(\overline{c_i^*} 2^i+c_i^* 2^{m+i})+2^{m+l}}\\
&\ +a^{\sum_{i\in D\setminus\{l\}}(\overline{c_i} 2^i+c_i 2^{m+i})+2^l}=A_{2^{m-l-1}-c'}
\end{split}\]
since
\[(\overline{c_{m-2}^*},\dots,\overline{c_l^*})=2^{m-l-1}-1-(c'-1)=2^{m-l-1}-c'\]
and
\[2^{m-l-1}-c'-1=(2^{m-l-1}-1)-c'=(\overline{c_{m-2}},\dots,\overline{c_l})\enspace.\]
Otherwise, if $m-1\in D$ then
\[\begin{split}
A_{c'}^{2^m}&=a^{2^{2m-1}+\sum_{i\in D\setminus\{l,m-1\}}(\overline{c_i^*} 2^i+c_i^* 2^{m+i})+2^{m+l}}\\
&\ +a^{2^{2m-1}+\sum_{i\in D\setminus\{l,m-1\}}(\overline{c_i} 2^i+c_i 2^{m+i})+2^l}\\
&=a^{2^{m-1}(2^m-1)}A_{2^{m-l-1}-c'}\enspace.
\end{split}\]

\item As a direct consequence we obtain that $A_{2^{m-l-2}}\in\ftwom$
    when $m-1\notin D$ and $a^{-2^{m-1}}A_{2^{m-l-2}}\in\ftwom$ when
    $m-1\in D$.

\item If $c'$ is odd then
\[A_{c'}=a^{\sum_{i\in D\setminus\{l\}\atop c_{m-1}=1}(c_i 2^i+\overline{c_i} 2^{m+i})}\big(a^{2^m}+a\big)^{2^l}\enspace.\]
\end{enumerate}

\begin{theorem}
Any Niho bent function in the univariate form, up to EA-equivalence, is
obtained as a sum of functions having the form of (\ref{eq:LK_coef}). Each
function making up the sum is defined by a monomial found in the
corresponding o-polynomial and has a particular set of nonzero coefficients
$A_i$ for $i=1,\dots,2^{r-1}$. Parameter $0<m-r<m$ is equal to the position
of the least significant one-digit in the binary expansion of the exponent in
this monomial.
\end{theorem}

\begin{IEEEproof}
By (\ref{eq:exH}), any Niho bent function in the bivariate form is equal to
$g(x,y)=\Tr_m\left(xF(yx^{2^m-2})+\mu y\right)$, where $F(z)$ defines an
o-polynomial over $\ftwom$. The linear term $\Tr_m(\mu y)$ can be dropped.

Polynomial $F(z)$ consists of power terms that can be treated separately
under the trace using Lemma~\ref{le:mon} and the results are added together.
Note that the identities $x=t+t^{2^m}$ and $y=at+a^{2^m}t^{2^m}$ used in
Lemma~\ref{le:mon} to obtain univariate formulas for functions in a bivariate
representation, assume a particular choice of a basis of $\ftwon$ as a
two-dimensional vector space over $\ftwom$. But we know that taking a
different basis just results in EA-equivalent functions.

Finally, by \cite[Result~1]{Ch88}, all terms in an o-polynomial have even
powers if $m>1$, i.e., $m-r=l$ from Lemma~\ref{le:mon} is not zero and $r<m$.
\end{IEEEproof}

From the result proven it follows that, up to EA-equivalence, the leading
term in a univariate polynomial giving a Niho bent function has degree
cyclotomic equivalent to $2^m+1$. This confirms the conjecture made in
\cite[Section~3]{DoLeCaCaFeGa06} for the particular case of bent binomials.
It also confirms that the only existing monomial Niho bent function is the
quadratic one $\Tr_m(at^{2^m+1})$ with $a\in\ftwom^*$. (need to check that
the coefficient is nonzero???)

Note that the function $g(x,y)=\Tr_m(\lambda x^{2^m-d}y^d)$ has algebraic
degree $m+wt(d)-wt(d-1)=m-l+1\leq m$ since $m-r=l>0$. Therefore, algebraic
degree of a Niho bent function is at most $m$ (as for any bent function).

\section{New Niho Bent Functions from Quadratic o-Monomials}
 \label{sec:new_qu}
In this section, we extend class (\ref{eq:LK1}) of bent functions for some
particular values of $m$ and $r$. This is done by inserting coefficients of
the power terms. These coefficients take just one of four possible values and
are repeated in the cycle of length $2^{c+1}$. Here we calculate the
corresponding function $F$ and later, selecting particular parameters, we
show that $F$ is an o-polynomial. This gives the proof of bentness.

For any integer $m>2$ take $n=2m$ and select $a\in\ftwon$ with $a+a^{2^m}=1$.
Take any $0\leq J<I<m-1$ and define
\[\begin{split}
A_1&=a^{2^I}+1\\
A_2&=a^{2^I}+a^{2^J}\\
A_3&=a^{2^I}+a^{2^J}+1\enspace.
\end{split}\]
Also fix integers $2<r\leq m$ and $0<c<r-1$ used to define the following
Boolean function over $\ftwon$
\begin{align}
 \label{eq:G_LK_qu}
&f(t)=\Tr_m\big(A_3 t^{2^{m-1}(2^m+1)}\big)\\
\nonumber &+\Tr_n\Bigg(\sum_{j=0}^{2^{r-c-2}-1}\bigg(A_1\sum_{i=1}^{2^c-1}t^{\left(2^{m-r}(2^{c+1}j+i)+1\right)(2^m-1)+1}\\
\nonumber &+A_2 t^{\left(2^{m-r}(2^{c+1}j+2^c)+1\right)(2^m-1)+1}\\
\nonumber &+A_1^{2^m}\sum_{i=2^c+1}^{2^{c+1}-1}t^{\left(2^{m-r}(2^{c+1}j+i)+1\right)(2^m-1)+1}\bigg)\\
\nonumber &+\sum_{j=0}^{2^{r-c-2}-2}A_3 t^{\left(2^{m-r}(2^{c+1}j+2^{c+1})+1\right)(2^m-1)+1}\Bigg)\enspace.
\end{align}
In the case when $r-c=2$ assume the last sum equal to zero.

It is easy to see that function (\ref{eq:G_LK_qu}) has the form of
(\ref{eq:LK_coef}) with coefficients repeated in a cycle of length $2^{c+1}$
as follows
\[\begin{split}
&\underbrace{i}_\text{$A_i=$}=\underbrace{1,\dots,{2^c-1}}_\text{$A_1$},\underbrace{2^c}_\text{$A_2$},\underbrace{2^c+1,\dots,{2^{c+1}-1}}_\text{$A_1^{2^m}$},
\underbrace{2^{c+1}}_\text{$A_3$},\\
&\dots,2^{r-1}\enspace.
\end{split}\]

Note that w.l.o.g. we can assume
\[A_1=a^{2^{I-J}}+1,\quad A_2=a^{2^{I-J}}+a,\quad A_3=a^{2^{I-J}}+a+1\]
since raising to the power $2^I$ $(I>0)$ permutes the set $\{a\in\ftwon\ |\
a+a^{2^m}=1\}$.

Further, note that $A_2,A_3\in\ftwom$ and $a+a^{2^m}=1$ implies that
$A_1^{2^m}+A_3=A_1+A_2$. Rewrite function $f(t)$ as
\[\begin{split}
&f(t)=\Tr_m\big(A_3 t^{2^{m-1}(2^m+1)}\big)\\
&\ +\Tr_n\Bigg(\sum_{j=0}^{2^{r-c-2}-1}t^{2^{m-r+c+1}j(2^m-1)+2^m}\\
&\ \times\bigg(A_1\sum_{i=1}^{2^c}t^{2^{m-r}i(2^m-1)}+A_1^{2^m}\sum_{i=2^c+1}^{2^{c+1}}t^{2^{m-r}i(2^m-1)}\\
&\ +(A_1+A_2)\big(t^{2^{m-r+c}(2^m-1)}+t^{2^{m-r+c+1}(2^m-1)}\big)\bigg)\Bigg)\\
&=\Tr_m\big(A_3 t^{2^{m-1}(2^m+1)}\big)+\Tr_n\Bigg(\frac{t^{2^{m-1}(2^m-1)}+1}{t^{2^{m-r+c+1}(2^m-1)}+1}t^{2^m}\\
&\ \times\bigg(A_1\frac{(t^{2^{m-r+c}(2^m-1)}+1)t^{2^{m-r}(2^m-1)}}{t^{2^{m-r}(2^m-1)}+1}\\
&\ +A_1^{2^m}\frac{(t^{2^{m-r+c}(2^m-1)}+1)t^{2^{m-r}(2^c+1)(2^m-1)}}{t^{2^{m-r}(2^m-1)}+1}\\
&\ +(A_1+A_2)\big(t^{2^{m-r+c}(2^m-1)}+t^{2^{m-r+c+1}(2^m-1)}\big)\bigg)\Bigg)\\
&=\Tr_m\big(A_3 t^{2^{m-1}(2^m+1)}\big)+\Tr_n\Bigg(\frac{t^{2^{m-1}(2^m+1)}+t^{2^m}}{(t^{2^m}+t)^{2^{m-r+c+1}}}\\
&\ \times\bigg((t^{2^m+1}+t^2)^{2^{m-r+c}}t^{2^{2m-r}}\frac{A_1+A_1^{2^m}t^{2^{m-r+c}(2^m-1)}}{(t^{2^m}+t)^{2^{m-r}}}\\
&\ +(A_1+A_2)(t^{2^m+1}+t^{2^{m+1}})^{2^{m-r+c}}\bigg)\Bigg)\enspace.
\end{split}\]
Here, in the case when $t^{2^m-1}=1$ we assume fractions are equal to zero.

Since $a\notin\ftwom$, the pair $(a,1)$ makes up a basis of $\ftwon$ as a
two-dimensional vector space over $\ftwom$. Then every element $t\in\ftwon$
can be uniquely represented as $ax+y$ with $(x,y)\in\ftwom\times\ftwom$.

Now if $x=0$ then $t=y$ and we obtain
\[f(y)=\Tr_m\big(A_3 y\big)\enspace.\]
For $x\neq 0$, denoting $s=a+y/x$ and since $s^{2^m}+s=1$, we obtain
\[\begin{split}
&f(ax+y)=\Tr_m\big(A_3 s^{2^{m-1}(2^m+1)}x\big)\\
&\ +\Tr_n\Bigg(x\frac{s^{2^{m-1}(2^m+1)}+s^{2^m}}{(s^{2^m}+s)^{2^{m-r+c+1}}}\\
&\ \times\bigg((s^{2^m+1}+s^2)^{2^{m-r+c}}s^{2^{2m-r}}\frac{A_1+A_1^{2^m}s^{2^{m-r+c}(2^m-1)}}{(s^{2^m}+s)^{2^{m-r}}}\\
&\ +(A_1+A_2)(s^{2^m+1}+s^{2^{m+1}})^{2^{m-r+c}}\bigg)\Bigg)\\
&=\Tr_m\big(A_3 s^{2^{m-1}(2^m+1)}x\big)+\Tr_n\Big(x(s^{2^{m-1}(2^m+1)}+s^{2^m})\\
&\times\big(s^{2^{2m-r}+2^{m-r+c}}(A_1+A_1^{2^m}s^{2^{m-r+c}(2^m-1)})\\
&+(A_1+A_2)(s+1)^{2^{m-r+c}}\big)\Big)\\
&=\Tr_m\big(A_3 s^{2^{m-1}(2^m+1)}x\big)\\
&+\Tr_n\Big(x\big(s^{2^{m-1}(2^m+1)}+s^{2^m}\big)\big((s^{2^{m-r}}+1)\\
&\times(s^{2^{m-r+c}}+A_1^{2^m})+(A_1+A_2)(s^{2^{m-r+c}}+1)\big)\Big)\\
&=\Tr_m\big(A_3 s^{2^{m-1}(2^m+1)}x\big)\\
&+\Tr_n\big(x(s^{2^{m-1}(2^m+1)}+s^{2^m})(s^{2^{m-r+c}+2^{m-r}}\\
&+(A_1+A_2+1)s^{2^{m-r+c}}+A_1^{2^m} s^{2^{m-r}}+A_3)\big)\\
&=\Tr_m\Big(x\big((s+1)(a^{2^I}+a^{2^J}+1)\\
&\ +s^{2^{m-r+c}+2^{m-r}}+a^{2^J}s^{2^{m-r+c}}+a^{2^I}s^{2^{m-r}}\big)\Big)\\
&=\Tr_m\Big(x\big(z^{2^{m-r+c}+2^{m-r}}+(a^{2^{m-r}}+a^{2^J})z^{2^{m-r+c}}\\
&+(a^{2^{m-r+c}}+a^{2^I})z^{2^{m-r}}+(a+z+1)(a^{2^I}+a^{2^J}+1)\\
&+a^{2^I+2^{m-r}}+a^{2^J+2^{m-r+c}}+a^{2^{m-r+c}+2^{m-r}}\big)\Big)\\
&=\Tr_m(xG(z))\enspace,
\end{split}\]
where $z=y/x$. Therefore, for any $x,y\in\ftwom$,
 \setlength{\arraycolsep}{0.14em}
\[f(ax+y)=\left\{\begin{array}{ll}
\Tr_m(xG(y/x)),&\quad\mbox{if}\quad x\neq 0\\
\Tr_m\big(A_3 y\big),&\quad\mbox{if}\quad x=0\enspace,
\end{array}\right.\]
 \setlength{\arraycolsep}{5pt}\noindent
and
\begin{align}
 \label{eq:F}
F(z)&=G(z)+A_3 z\\
\nonumber &=z^{2^{m-r+c}+2^{m-r}}+(a^{2^{m-r}}+a^{2^J})z^{2^{m-r+c}}\\
\nonumber &+(a^{2^{m-r+c}}+a^{2^I})z^{2^{m-r}}+(a+1)(a^{2^I}+a^{2^J}+1)\\
\nonumber &+a^{2^I+2^{m-r}}+a^{2^J+2^{m-r+c}}+a^{2^{m-r+c}+2^{m-r}}\enspace.
\end{align}

Note that
\[0\leq m-r<m-2\quad\mbox{and}\quad m-r<m-r+c<m-1\enspace.\]
In particular, taking $J=m-r$ and $I=m-r+c$ we obtain
\[F(z)=z^{2^I+2^J}+\mbox{const}\enspace.\]
The full range is $0\leq J<I<m$. So we have to consider separately the case
when $I=m-1$. If $J=m-2$ then $F(z)=z^{2^{m-1}+2^{m-2}}$ and applying
transformation $zF(z^{-1})$ we obtain $z^{2^{m-2}}$ that is a Frobenius
o-polynomial if and only if $m$ is odd. Transformation $zF(z^{-1})$ of
o-polynomials translated in terms of the associated bent functions results in
a particular case of EA-equivalence (see 3.1.2 in \cite{CaMe11}). Therefore,
the quadratic o-polynomial listed under number~\ref{it:5} corresponds to the
Niho bent function that is EA-equivalent to the function obtained from the
Frobenius mapping $z^{2^{m-2}}$ with $m$ odd.

For any integer $m>1$ take $n=2m$ and select $a\in\ftwon$ with $a+a^{2^m}=1$.
Take any $0\leq J<m-1$ and define $r=m-J$,
\[A_1=a^{2^{m-1}}\quad\mbox{and}\quad A_3=a^{2^{m-1}}+a^{2^J}\]
for the following Boolean function over $\ftwon$
\begin{align}
 \label{eq:G_LK2}
f(t)=&\Tr_m\big(A_3 t^{2^{m-1}(2^m+1)}\big)\\
\nonumber&+\Tr_n\Bigg(A_1\sum_{i=1}^{2^{r-1}-1}t^{(2^{m-r}i+1)(2^m-1)+1}\Bigg)\enspace.
\end{align}



\subsection{Bent Functions with $2^{m-2}$ Niho Exponents}
 \label{ssec:1}
Assume $m>3$ is odd and take function (\ref{eq:G_LK_qu}) with $r=m-1$, $c=1$,
$I=2$ and $J=1$. Then, by (\ref{eq:F}),
\[F(z)=z^6+a^6+(a+1)(a^4+a^2+1)\]
and, ignoring the constant term, this is an o-polynomial $z^6$. Therefore,
function (\ref{eq:G_LK_qu}) with such parameters is a bent function.

\begin{note}
This bent function can also have a more general form when taking any
$a\in\ftwon$ with $a+a^{2^m}\neq 0$. Define coefficients differently as
\[\begin{split}
A_1&=a^{6\cdot 2^m}+a^{2^{m+2}+2}\\
A_2&=a^{2^{m+2}+2}+a^{2^{m+1}+4}\\
A_3&=a^6+a^{6\cdot 2^m}\enspace.
\end{split}\]
Obviously, if $a+a^{2^m}=1$ then these coefficients are the same as defined
originally in this section. Still, this extension does not contain any new
bent functions, up to EA-equivalence. Indeed, let $a+a^{2^m}=b\in\ftwom$ and
substitute $t$ in the new function for $b^{-6}t$. This results in a similar
function except for $a/b$ taken instead of $a$. But $a/b+(a/b)^{2^m}=1$ as we
assumed originally.
\end{note}

\begin{note}
For $m=3$, take any $a\in\F_{2^6}$ with $a+a^8\neq 0$. Then for the basis
$(a,1)$, o-polynomial $z^6$ corresponds to the bent function that, up to the
addition of a linear term, has the following form
\[f(t)=\Tr_6(a^{36}t^{36})+\Tr_6(a^{22}t^{22})\enspace.\]
Substituting $at$ for $t$ we obtain the EA-equivalent bent function
$\Tr_3(t^9)+\Tr_6(t^{22})$ that is exactly function (\ref{eq:LK1}) with
$r=m-1=2$ and $a+a^{2^m}=1$.
\end{note}

\subsection{Bent Functions with $2^{m-k-1}$ Niho Exponents}
Assume $m=4k-1>3$ and take function (\ref{eq:G_LK_qu}) with $r=3k-1$, $c=k$,
$I=2k$ and $J=k$. Then, by (\ref{eq:F}),
\[F(z)=z^{2^{2k}+2^k}+a^{2^{2k}+2^k}+(a+1)(a^{2^{2k}}+a^{2^k}+1)\]
and, ignoring the constant term, this is an o-polynomial $z^{2^{2k}+2^k}$.
Therefore, function (\ref{eq:G_LK_qu}) with such parameters is a bent
function.

\begin{note}
This bent function can also have a more general form when taking any
$a\in\ftwon$ with $a+a^{2^m}\neq 0$. Define coefficients differently as
\[\begin{split}
A_1&=a^{2^{m+2k}+2^{m+k}}+a^{2^{m+2k}+2^k}\\
A_2&=a^{2^{m+2k}+2^k}+a^{2^{2k}+2^{m+k}}\\
A_3&=a^{2^{m+2k}+2^{m+k}}+a^{2^{2k}+2^k}\enspace.
\end{split}\]
Obviously, if $a+a^{2^m}=1$ then these coefficients are the same as defined
originally in this section. Still, this extension does not contain any new
bent functions, up to EA-equivalence. Indeed, let $a+a^{2^m}=b\in\ftwom$ and
substitute $t$ in the new function for $b^{-(2^{2k}+2^k)}t$. This results in
a similar function except for $a/b$ taken instead of $a$. But
$a/b+(a/b)^{2^m}=1$ as we assumed originally.
\end{note}

\begin{note}
For $k=1$, o-polynomial $z^{2^{2k}+2^k}=z^6$ is of Segre.
\end{note}

\subsection{Bent Functions with $2^{m-2k-2}$ Niho Exponents}
Assume $m=4k+1>5$ and take function (\ref{eq:G_LK_qu}) with $r=2k$, $c=k$,
$I=3k+1$ and $J=2k+1$. Then, by (\ref{eq:F}),
\[\begin{split}
F(z)=z^{2^{3k+1}+2^{2k+1}}&+a^{2^{3k+1}+2^{2k+1}}\\
&+(a+1)(a^{2^{3k+1}}+a^{2^{2k+1}}+1)
\end{split}\]
and, ignoring the constant term, this is an o-polynomial
$z^{2^{3k+1}+2^{2k+1}}$. Therefore, function (\ref{eq:G_LK_qu}) with such
parameters is a bent function.

\begin{note}
This bent function can also have a more general form when taking any
$a\in\ftwon$ with $a+a^{2^m}\neq 0$. Define coefficients differently as
\[\begin{split}
A_1&=a^{2^{m+3k+1}+2^{m+2k+1}}+a^{2^{m+3k+1}+2^{2k+1}}\\
A_2&=a^{2^{m+3k+1}+2^{2k+1}}+a^{2^{3k+1}+2^{m+2k+1}}\\
A_3&=a^{2^{3k+1}+2^{2k+1}}+a^{2^{m+3k+1}+2^{m+2k+1}}\enspace.
\end{split}\]
Obviously, if $a+a^{2^m}=1$ then these coefficients are the same as defined
originally in this section. Still, this extension does not contain any new
bent functions, up to EA-equivalence. Indeed, let $a+a^{2^m}=b\in\ftwom$ and
substitute $t$ in the new function for $b^{-(2^{3k+1}+2^{2k+1})}t$. This
results in a similar function except for $a/b$ taken instead of $a$. But
$a/b+(a/b)^{2^m}=1$ as we assumed originally.
\end{note}

\begin{note}
For $k=1$ (i.e., $m=5$), o-polynomial $z^{24}$ is obtained from the Frobenius
mapping $z^8$ by transformation $zF(z^{-1})$ that preserves equivalence of
o-polynomials and EA-equivalence of the corresponding bent functions (see
3.1.2 in \cite{CaMe11}).
\end{note}

\subsection{Bent Functions with $2^{m-2}$ Niho Exponents}
Assume $m=2k-1>3$ and take function (\ref{eq:G_LK_qu}) with $r=m-1$, $c=k-1$,
$I=k$ and $J=1$. Then, by (\ref{eq:F}),
\[F(z)=z^{2^k+2}+a^{2^k+2}+(a+1)(a^{2^k}+a^2+1)\]
and, ignoring the constant term, this is an o-polynomial $z^{2^k+2}$.
Therefore, function (\ref{eq:G_LK_qu}) with such parameters is a bent
function.

\begin{note}
Note that
\[(2^k+2)(1-2^{k-1})\equiv 1\pmod{2^m-1}\]
which means that the inverse of $z^{2^k+2}$ is $z^{1-2^{k-1}}$. The latter is
obtained from the Frobenius o-polynomial $z^{2^{k-1}}$ by transformation
$zF(z^{-1})$ that preserves equivalence of o-polynomials (see 3.1.2 in
\cite{CaMe11}). Since the inverse of an o-polynomial is an o-polynomial we
conclude that $z^{2^{k-1}}$ and $z^{2^k+2}$ are equivalent o-polynomials.

Transformation $zF(z^{-1})$ of o-polynomials translated in terms of the
associated bent functions results in a particular case of EA-equivalence. On
the contrary, inverse o-polynomial does not correspond to the EA-equivalent
bent function. This illustrates the case when two EA-inequivalent Niho bent
functions arise from equivalent o-polynomials.
\end{note}

\begin{note}
For $k=2$, o-polynomial $z^{2^k+2}=z^6$ is of Segre and for $k=1$,
$z^{2^k+2}=z^4$ is Frobenius mapping.
\end{note}

\begin{table*}
\caption{Niho bent functions from equivalent o-monomials}
 \label{ta:1}
\renewcommand{\arraystretch}{1.9}
\centering
\begin{minipage}{16cm}
\begin{tabular}{c|lc|lc|lc}
\hline $m$&$G_1(z)$&$d_1$&$G_2(z)$&$d_2$&$G_3(z)$&$d_3$\\
\hline $2k-1$&$2^k$&$k$&$2^{k-1}$&$k+1$&$2^k+2$&$m$\\
\hline $4k+1$&\multirow{2}{*}{$6$}&\multirow{2}{*}{$m$}&\multirow{2}{*}{$\sum_{i=0}^{\frac{m-3}{2}}2^{2i+1}+2^{m-1}$}&\multirow{2}{*}{$m$}&$2+\sum_{i=1}^k 2^{4i}+\sum_{i=1}^k 2^{4i-1}$&$m$\\
       $4k+3$&&&&&$4+\sum_{i=1}^k 2^{4i}+\sum_{i=1}^k 2^{4i+1}$&$m-1$\\
\hline \multirow{2}{*}{$4k-1$}&\multirow{2}{*}{$2^{2k}+2^k$}&\multirow{2}{*}{$3k$}&\multirow{2}{*}{$2^m-2^{3k-1}+2^{2k}-2^k$}&\multirow{2}{*}{$3k$}&$2+\sum_{i=1}^{\frac{k-1}{2}} 2^{2i}+\sum_{i=\frac{k-1}{2}}^{\frac{3k-3}{2}}2^{2i+1}$&$m$\footnote{$k>1$ odd}\\
       &&&&&$2^k+\sum_{i=\frac{k}{2}}^{\frac{3k-2}{2}} 2^{2i+1}+\sum_{i=\frac{3k}{2}}^{2k-1}2^{2i}$&$3k$\footnote{$k>0$ even}\\
\hline \multirow{2}{*}{$4k+1$}&\multirow{2}{*}{$2^{3k+1}+2^{2k+1}$}&\multirow{2}{*}{$2k+1$}&\multirow{2}{*}{$2^m-2^{3k+1}+2^{2k+1}-2^k$}&\multirow{2}{*}{$3k+2$}&$2^{k+1}+\sum_{i=\frac{k+1}{2}}^{\frac{3k-1}{2}}2^{2i+1}+\sum_{i=\frac{3k+1}{2}}^{2k}2^{2i}$&$3k+1$\footnote{$k$ odd}\\
       &&&&&$2+\sum_{i=1}^{\frac{k}{2}}2^{2i}+\sum_{i=\frac{k}{2}}^{\frac{3k-2}{2}}2^{2i+1}$&$m^\dagger$\\
\hline \multirow{2}{*}{$2k-1$}&\multirow{2}{*}{$3\cdot 2^k+4$}&\multirow{2}{*}{$m-1$}&\multirow{2}{*}{$3\cdot 2^{k-1}-2$}&\multirow{2}{*}{$m$}&$2^k+\sum_{i=\frac{k+1}{2}}^{k-1}2^{2i}$&$k^\ast$\\
       &&&&&$2+\sum_{i=1}^{\frac{k-2}{2}}2^{2i}$&$m$\footnote{$k>2$ even}\\
\hline
\end{tabular}
\end{minipage}
\end{table*}

\begin{table*}
\caption{Niho bent functions from equivalent o-polynomials}
 \label{ta:2}
\renewcommand{\arraystretch}{1.9}
\centering
\begin{minipage}{16cm}
\begin{tabular}{c|lc|lc|lc}
\hline $m$&$G_1(z)$&$d_1$&$G_2(z)$&$d_2$&$G_3(z)$&$d_3$\\
\hline $2k-1$&$z^{2^k}+z^{2^k+2}+z^{3\cdot 2^k+4}$&$m$\footnote{$k>2$}&$z(z^{2^k+1}+z^3+z)^{2^{k-1}-1}$&$m$&&\\
\hline odd&$z^{\frac{1}{6}}+z^{\frac{1}{2}}+z^{\frac{5}{6}}$&$m$&&&$zG_2(z^{-1})$&\\
\hline
\end{tabular}
\end{minipage}
\end{table*}

\section{New Niho Bent Functions from the Cubic o-Monomial}
 \label{sec:new_cu}
In this section, we extend class (\ref{eq:LK1}) of bent functions for any odd
$m$ and $r=m-2$. This is done by inserting coefficients of the power terms.
These coefficients take just one of eight possible values and are repeated in
the cycle of length $2^{(m+1)/2}$. Here we calculate the corresponding
function $F$ and showing that $F$ is an o-polynomial, we give the proof of
bentness.

For any integer $m$ take $n=2m$ and select $a\in\ftwon$ with $a+a^{2^m}=1$.
Take any $0<J+1<I<m-1$ and define
\[\begin{split}
A_1&=a^{3\cdot 2^{I-1}}\\
A_2&=a^{2^I}(a^{2^{I-1}}+a^{2^J})\\
A_3&=a^{3\cdot 2^{I-1}+2^J}+(a+1)^{3\cdot 2^{I-1}+2^J}\enspace.
\end{split}\]

Also define the following Boolean function over $\ftwon$
\begin{align}
 \label{eq:G_LK_cu}
&f(t)=\Tr_m\big(A_3t^{2^{m-1}(2^m+1)}\big)\\
\nonumber &+\Tr_n\Bigg(\sum_{l=0}^{2^{m-I-2}-1}\bigg(A_1\sum_{j=0}^3 a^{j2^{I-1}(2^m-1)}\sum_{i=1}^{2^{I-J-1}-1}t^{\left(2^J(2^{I-J-1}(4l+j)+i)+1\right)(2^m-1)+1}\\
\nonumber &+A_2\sum_{j=0}^2 a^{j2^{I-1}(2^m-1)}t^{\left(2^J(2^{I-J-1}(4l+j)+2^{I-J-1})+1\right)(2^m-1)+1}\bigg)\\
\nonumber &+A_3\sum_{l=0}^{2^{m-I-2}-2}t^{\left(2^J(2^{I-J-1}(4l+3)+2^{I-J-1})+1\right)(2^m-1)+1}\Bigg)\enspace.
\end{align}
In the case when $I=m-2$ assume the last sum is equal to zero.

It is easy to see that function (\ref{eq:G_LK_cu}) has the form of
(\ref{eq:LK_coef}) with coefficients repeated in a cycle of length $2^{c+1}$
(with $c=I-J$ and where $e=2^{I-1}(2^m-1)$) as follows
\[\begin{split}
&\underbrace{i}_\text{$A_i=$}=\underbrace{1,\dots,{2^{c-1}-1}}_\text{$A_1$},\underbrace{2^{c-1}}_\text{$A_2$},\underbrace{2^{c-1}+1,\dots,{2^c-1}}_\text{$a^e A_1$},
\underbrace{2^c}_\text{$a^e A_2$},\\
&\underbrace{2^c+1,\dots,{3\cdot 2^{c-1}-1}}_\text{$a^{2e}A_1=(a^e A_1)^{2^m}$},\underbrace{3\cdot 2^{c-1}}_\text{$a^{2e}A_2=A_2^{2^m}$},\\
&\underbrace{3\cdot 2^{c-1}+1,\dots,{2^{c+1}-1}}_\text{$a^{3e}A_1=A_1^{2^m}$},\underbrace{2^{c+1}}_\text{$A_3$},\dots,2^{m-J-1}\enspace.
\end{split}\]
Note that $a^e A_2\in\ftwom$.

Further, note that
\[\begin{split}
&A_2 a^{3\cdot 2^{I-1}(2^m-1)}+A_3\\
&=(a+1)^{3\cdot 2^{I-1}}+a^{2^{I-1}(3\cdot 2^m-1)+2^J}+a^{3\cdot 2^{I-1}+2^J}+(a+1)^{3\cdot 2^{I-1}+2^J}\\
&=a^{2^J}(a+1)^{3\cdot 2^{I-1}}+a^{2^{I-1}(3\cdot 2^m-1)+2^J}+a^{3\cdot 2^{I-1}+2^J}\\
&=a^{2^J}(a^{2^I}+a^{2^{I-1}}+1)+a^{2^{I-1}(3\cdot 2^m-1)+2^J}\\
&=a^{2^J-2^{I-1}}\big(a^{3\cdot 2^{I-1}}+a^{2^I}+a^{2^{I-1}}+(a+1)^{3\cdot 2^{I-1}}\big)\\
&=a^{2^J-2^{I-1}}
\end{split}\]
and rewrite function $f(t)$ as
\[\begin{split}
&f(t)=\Tr_n\Bigg(a^{3\cdot 2^{I-1}+2^J}t^{2^{m-1}(2^m+1)}\\
&+\sum_{l=0}^{2^{m-I-2}-1}\bigg(A_1\sum_{i=1}^{2^{I-J-1}}t^{2^J(2^{I-J+1}l+i)(2^m-1)+2^m}\sum_{j=0}^3 (at)^{j2^{I-1}(2^m-1)}\\
&+(A_1+A_2)t^{2^{I-1}(4l+1)(2^m-1)+2^m}\sum_{j=0}^3 (at)^{j2^{I-1}(2^m-1)}\\
&+\big(A_2 a^{3\cdot 2^{I-1}(2^m-1)}+A_3\big)t^{2^{I+1}(l+1)(2^m-1)+2^m}\bigg)\Bigg)\\
&=\Tr_n\bigg(a^{3\cdot 2^{I-1}+2^J}t^{2^{m-1}(2^m+1)}\\
&+t^{2^{m+J}}\frac{\big(t^{2^{m-1}(2^m+1)}+t^{2^m}\big)\big(t^{2^m}+t\big)^{2^{I-1}}\big((at)^{2^m}+at\big)^{2^{I+1}}}{\big(t^{2^m}+t\big)^{2^{I+1}+2^J}\big((at)^{2^m}+at\big)^{2^{I-1}}}\\
&+a^{2^J-2^{I-1}}t^{2^{m+I-1}}\frac{\big(t^{2^{m-1}(2^m+1)}+t^{2^m}\big)\big((at)^{2^m}+at\big)^{2^{I+1}}}{\big(t^{2^m}+t\big)^{2^{I+1}}\big((at)^{2^m}+at\big)^{2^{I-1}}}\\
&+a^{2^J-2^{I-1}}t^{2^{m+I+1}}\frac{t^{2^{m-1}(2^m+1)}+t^{2^m}}{(t^{2^m}+t)^{2^{I+1}}}\bigg)\\
&=\Tr_n\Bigg(a^{3\cdot 2^{I-1}+2^J}t^{2^{m-1}(2^m+1)}\\
&+t^{2^{m+J}}\frac{\big(t^{2^{m-1}(2^m+1)}+t^{2^m}\big)\big((at)^{2^m}+at\big)^{3\cdot 2^{I-1}}}{\big(t^{2^m}+t\big)^{3\cdot 2^{I-1}+2^J}}\\
&+a^{2^J-2^{I-1}}t^{2^{m+I-1}}\frac{\big(t^{2^{m-1}(2^m+1)}+t^{2^m}\big)\big((at)^{2^m}+at\big)^{3\cdot 2^{I-1}}}{\big(t^{2^m}+t\big)^{2^{I+1}}}\\
&+a^{2^J-2^{I-1}}t^{2^{m+I+1}}\frac{t^{2^{m-1}(2^m+1)}+t^{2^m}}{(t^{2^m}+t)^{2^{I+1}}}\Bigg)\enspace.
\end{split}\]
Here, in the case when $t^{2^m-1}=1$ or $(at)^{2^m-1}=1$ we assume the
relevant fractions are equal to zero.

Since $a\notin\ftwom$, the pair $(a+1,1)$ makes up a basis of $\ftwon$ as a
two-dimensional vector space over $\ftwom$. Then every element $t\in\ftwon$
can be uniquely represented as $(a+1)x+y$ with $(x,y)\in\ftwom\times\ftwom$.

Now if $x=0$ then $t=y$ and we obtain
\[f(y)=\Tr_m\big(A_3 y\big)\enspace.\]
For $x\neq 0$, denoting $s=a+1+y/x$ and since $s^{2^m}+s=1$, we obtain
\[\begin{split}
&f((a+1)x+y)=\Tr_n\Bigg(a^{3\cdot 2^{I-1}+2^J}s^{2^{m-1}(2^m+1)}x\\
&+s^{2^{m+J}}x\frac{\big(s^{2^{m-1}(2^m+1)}+s^{2^m}\big)\big((as)^{2^m}+as\big)^{3\cdot 2^{I-1}}}{\big(s^{2^m}+s\big)^{3\cdot 2^{I-1}+2^J}}\\
&+a^{2^J-2^{I-1}}s^{2^{m+I-1}}x\frac{\big(s^{2^{m-1}(2^m+1)}+s^{2^m}\big)\big((as)^{2^m}+as\big)^{3\cdot 2^{I-1}}}{\big(s^{2^m}+s\big)^{2^{I+1}}}\\
&+a^{2^J-2^{I-1}}s^{2^{m+I+1}}x\frac{s^{2^{m-1}(2^m+1)}+s^{2^m}}{(s^{2^m}+s)^{2^{I+1}}}\Bigg)\\
&=\Tr_n\Big(a^{3\cdot 2^{I-1}+2^J}s^{2^{m-1}(2^m+1)}x\\
&+x\big(s^{2^{m+J}}+a^{2^J-2^{I-1}}s^{2^{m+I-1}}\big)\big(s^{2^{m-1}(2^m+1)}+s^{2^m}\big)(a+s+1)^{3\cdot 2^{I-1}}\\
&+a^{2^J-2^{I-1}}s^{2^{m+I+1}}x\big(s^{2^{m-1}(2^m+1)}+s^{2^m}\big)\Big)\\
&=\Tr_n\Big(a^{3\cdot 2^{I-1}+2^J}s^{2^{m-1}(2^m+1)}x\\
&+x\big(z^{3\cdot 2^{I-1}+2^J}+a^{3\cdot 2^{I-1}+2^J}\big)\big(s^{2^{m-1}(2^m+1)}+z+a\big)\Big)\\
&=\Tr_m\Big(x\big(z^{3\cdot 2^{I-1}+2^J}+a^{3\cdot 2^{I-1}+2^J}\big)(z+a)\\
&+x\big(z^{3\cdot 2^{I-1}+2^J}+(a+1)^{3\cdot 2^{I-1}+2^J}\big)(z+a+1)\Big)\\
&=\Tr_m\Big(x\big(a^{3\cdot 2^{I-1}+2^J}+(a+1)^{3\cdot 2^{I-1}+2^J}\big)(z+a)\\
&+x\big(z^{3\cdot 2^{I-1}+2^J}+(a+1)^{3\cdot 2^{I-1}+2^J}\big)\Big)\\
&=\Tr_m(xG(z))\enspace,
\end{split}\]
where $z=y/x$. Therefore, for any $x,y\in\ftwom$,
 \setlength{\arraycolsep}{0.14em}
\[f((a+1)x+y)=\left\{\begin{array}{ll}
\Tr_m(xG(y/x)),&\quad\mbox{if}\quad x\neq 0\\
\Tr_m\big(A_3 y\big),&\quad\mbox{if}\quad x=0\enspace,
\end{array}\right.\]
 \setlength{\arraycolsep}{5pt}\noindent
and
\[\begin{split}
&F(z)=G(z)+A_3 z\\
&=a\big(a^{3\cdot 2^{I-1}+2^J}+(a+1)^{3\cdot 2^{I-1}+2^J}\big)\\
&+z^{3\cdot 2^{I-1}+2^J}+(a+1)^{3\cdot 2^{I-1}+2^J}\\
&=z^{3\cdot 2^{I-1}+2^J}+a^{3\cdot 2^{I-1}+2^J+1}+(a+1)^{3\cdot 2^{I-1}+2^J+1}\enspace.
\end{split}\]

In particular, take $m=2k-1>5$ and $I=k+1$, $J=2$. Then
\[F(z)=z^{3\cdot 2^k+4}+a^{3\cdot 2^k+5}+(a+1)^{3\cdot 2^k+5}\]
and, ignoring the constant term, this is an o-polynomial $z^{3\cdot 2^k+4}$.
Therefore, function (\ref{eq:G_LK_cu}) with such parameters is a bent
function.

\begin{note}
For $k=2$ (i.e., $m=3$), o-polynomial $z^{3\cdot 2^k+4}=z^{16}=z^2$ is
Frobenius mapping. For $k=3$ (i.e., $m=5$), o-polynomial $z^{3\cdot
2^k+4}=z^{28}$ is obtained from the Frobenius mapping $z^4$ by transformation
$zF(z^{-1})$ that preserves equivalence of o-polynomials and EA-equivalence
of the corresponding bent functions (see 3.1.2 in \cite{CaMe11}).
\end{note}

Now it is easy to find a Niho bent function that corresponds to the following
o-trinomial of degree three
\[F(z)=z^{2^k}+z^{2^k+2}+z^{3\cdot 2^k+4}\quad\mbox{with}\quad m=2k-1\enspace.\]
Assume $n=2m$ with $m=2k-1>5$ and select $a\in\ftwon$ with $a+a^{2^m}=1$.
Since $F(z)$ is a sum of three o-monomials, we need to take a sum of three
Niho bent functions that correspond to each of them. The first linear term is
a Frobenius map that gives bent function (\ref{eq:LK1}) with $r=k-1$. The
second term is quadratic and corresponds to bent function (\ref{eq:G_LK_qu})
taken with $r=m-1$, $c=k-1$, $I=k$ and $J=1$. Finally, the third term of
degree three corresponds to the bent function (\ref{eq:G_LK_cu}) (because of
a differently chosen basis in this case we need to take $a+1$ in stead of $a$
in coefficients $A_1$, $A_2$, $A_3$). Added together, the resulting bent
function has the form of (\ref{eq:LK1}) with $r=m-1$ and coefficients of
power terms taking on one of at most ten different values.

\section{Conclusions}
From our main results in Section~\ref{sec:new_qu} it follows that for any odd
$m>5$ there exist three (two for $m=5$) classes of Niho bent functions that
have the form of (\ref{eq:G_LK_qu}). These functions correspond to quadratic
o-monomials. Up to EA-equivalence, these cases cover all the existing
quadratic o-monomials.

In Table~\ref{ta:1}, we present exponents for o-monomials $G_i(z)$, where
$G_2(z)=G_1^{-1}(z)$ and $G_3(z)=(zG_2(z^{-1}))^{-1}$; $d_i$ is the algebraic
degree of a Niho bent function obtained from $G_i(z)$. Explicit expressions
for $G_3(z)$ can be verified directly using formulas found in
\cite[Chap~5]{Gl83}. Since two EA-equivalent functions have the same
algebraic degree, one can easily make conclusions on EA-inequivalence of many
of the Niho bent functions arising from quadratic o-monomials using data in
Table~\ref{ta:1}. We also checked with a computer that for $m=5$, the Niho
bent function corresponding to $z^{2^k+2}$ is EA-inequivalent to any of the
Niho function of degree $m$ contained in (2)-(3) of
Subsection~\ref{sssec:Niho} and to any function arising from $G_1$, $G_2$,
and $G_3$ with $G_1(z)=z^6$. Further, for $m=5$ and $G_1(z)=z^6$ we got that
the Niho bent functions arising from $G_1$, $G_2$, and $G_3$ are mutually
EA-equivalent but they are EA-inequivalent to any Niho function of degree $m$
contained in (2)-(3) of Subsection~\ref{sssec:Niho}.



\end{document}